\documentclass[aps,twocolumn,preprintnumbers,amsmath,amssymb,floatfix,showkeys]{revtex4}
\usepackage{graphicx}
\usepackage{amsmath,amssymb,latexsym}
\usepackage{bm}

\newcommand{\bea}{\begin{eqnarray}}
\newcommand{\eea}{\end{eqnarray}}

\begin{document}

\title{Spacing Characterization in Al-Cu Alloys Directionally Solidified Under Transient Growth Conditions}

\author{Morteza Amoorezaei}
\author{Sebastian Gurevich}
\author{Nikolas Provatas}

\affiliation{Department of Materials Science and Engineering,\\ McMaster University, Hamilton, ON, L8S4L7}

\date{\today}

\begin{abstract}
We study spacing selection in directional solidification of Al-Cu alloys under transient growth conditions. New experimental results are presented which reveal that dendritic spacing versus solidification rate evolves in an almost 
step-wise fashion, consistent with previous theoretical predictions of Langer and co-workers. Phase field simulations of directional solidification with dynamical growth conditions approximating those in the experiments confirm this behavior. Changes in dendrite arm spacing is shown to be consistent with dendrite instabilities confined, initially, to sub-domains, rather than the entire system. This is due to the rapid variation in growth conditions, which prevent the system from adapting as a whole but, rather, in a succession of quasi-isolated domains.
\end{abstract}

\maketitle

\section{Introduction}

Solidification microstructure is the starting point of any 
casting operation. Dendritic spacing and morphology established 
during casting often sets the scales of the downstream 
microstructure during manufacturing of alloys. This is particularly 
true in emerging technologies such as twin belt casting, where a 
reduced amount of thermomechanical downstream processing 
reduces the possibility of modifying microstructure length scales 
from that determined at the time of solidification.

Predicting columnar microstructure in cast alloys has been traditionally 
studied in the context of Bridgeman type directional solidification conditions. 
Most studies have focused on the problem of primary and secondary arm 
spacing in dendrite arrays of thin liquid films of organic alloys, directionally 
solidified under steady-state cooling conditions, i.e. a sample is 
pulled through a constant thermal gradient at a constant pulling speed. 
Careful experiments on steady state directional solidification reveal a 
reproducible correlation between spacing and pulling speed 
\cite{Trivedi:80,Kirkaldy:96}.
Studies of steady state directional solidification have developed so-called 
geometric models to relate spacing to solidification processing parameters 
such as the pulling speed $V$, thermal gradient $G$ and the alloy 
concentration $c_o$.  

In geometric theories of spacing selection  the structure and 
mathematical form of the dendrite arms is first assumed and a 
subsequent consistency relation is derived for the arm spacing (also 
referred to as ``wavelength" in the literature). The construction and 
assumptions that went into setting up the geometry of the dendrite 
array lead to at least one phenomenological  parameter that is then 
fit to match the theory onto specific experimental spacing selection 
data \cite{Trivedi:83,Kirkaldy:97,Hunt:79}. While 
useful in helping to elucidate some aspects of spacing selection, such 
theories lack the fundamental element of microstructure predictability: 
the ability to self-consistently generate the morphology of the structure 
they are trying to predict anything about. It is also not clear how such 
theories hold up to a change of conditions away from those of the  
experiments they were constructed to model.

While steady state directional solidification is an important academic 
paradigm, it is not a realistic representation of the conditions prevalent 
during industrial casting, which normally occurs under rapidly changing  
growth conditions. Near a chill surface or, indeed, throughout 
the entire sample thickness in the case of very thin strips, the thermal 
gradient and solidification speed are neither constant nor independent 
of each other. Recognizing this, several research groups have attempted 
to extend geometrical models to include unsteady state processing 
conditions. For example, Garcia and co-workers \cite{Garcia:08} and 
Kirkaldy and co-workers \cite{Kirkaldy:97}  have 
attempted to link the spacing of dendrite arms to the cooling rate 
$\dot{q}$. Once again, as with any geometric theory, phenomenological 
parameters are introduced to fit the model with experiments, although 
in this case the fits are not as good as in the steady state case. 

The importance of transient thermal history processing conditions in 
establishing as-cast microstructures has been apparent since the 
theoretical work of Warren and  Langer \cite{Langer:90,Langer:93} 
and the experimental studies of  Losert and Huang \cite{Huang:99,Losert:96} 
on alloys of succinonitrile (SCN). Warren and Langer performed 
an analysis on the stability of dendritic arrays \cite{Langer:93,Langer:90} 
and showed that they remain stable to period doubling instability 
over a range of pulling speeds, contrary to the predictions of any 
geometric theory, steady state or transient.  Their work suggested 
a band of available spacings rather than a unique selection. In turn, 
Huang et al \cite{Huang:99} showed that by changing the rate of the 
pulling speed it is possible to obtain different spacings for a given set of final 
growth conditions. Moreover, Losert et al \cite{Losert:96}  observed that 
under a more gradual change in pulling speed the spacing still remained stable 
over a range of pulling speeds,  consistent with the Warren and Lager predictions. 
They also noted the presence of a sharp increase in spacing which, consistent with 
the predictions of Warren and Langer, could be attributed to period doubling.

The results in the literature seem to point to two extremes. Under steady state 
conditions, i.e. $dV/dt=0$, the dendrite arm spacing appears to scale as a 
power-law of the pulling speed $V$, a result at least borne out qualitatively 
by geometrical models. On the other hand, under transient solidification  
conditions, dendrite spacing and structure  seem to depend strongly on 
transient history and initial conditions, at least in the idealized setting of a linear 
stability analysis or for well controlled experimental SCN dendrite arrays. 
The lack of  unified theory to explain both these regimes likely points to an incomplete 
picture of a fundamental physics underlying microstructure selection in 
solidification. It also points to the need for a robust  theory and modeling 
formalism that predicts the evolution of dendritic morphologies and growth rate 
{\it self-consistently}, as function of only the input material parameters and 
cooling conditions, steady state or transient. 

Phase field theory has emerged in recent years as promising candidate of 
a fundamental and self-consistent theory for modeling solidification 
microstructures. The first simulations to test spacing versus pulling 
speed in alloys date back to the work of  Warren and Boettinger 
\cite{Boettinger:00}, who found a monotonic band of spacings versus pulling 
speed. The small system size used, however, precluded a quantitative comparison 
with experiments. Nowadays the phase field methodology has become more 
quantitative by ``marrying"  simulations of phase field models in the so-called 
{\it thin interface limit} \cite{Karma:01, Echebarria:04} with novel simulation 
techniques like adaptive mesh refinement \cite{Provatas:98}.  A first step using 
phase field models to quantitatively model spacing in directional solidification 
was taken by Greenwood et. al  \cite{Greenwood:04} in 2D and Dantzig and 
co-workers in 3D \cite{review:05}. These works modeled steady state 
directional solidification in SCN alloys and found very good agreement (in the 2D 
limit) with the 2D steady state spacing experiments. These studies suggested 
that, at least under steady state (i.e. Bridgeman growth) conditions and one 
type of initial condition (morphologically noisy initial interface), there could be a single 
crossover scaling function interpolating between the two power-law spacing regimes 
seen experimentally and modeled semi-empirically by geometrical models.

Despite the success of phase field modeling in predicting steady state spacing, 
as well as other steady state properties such as cell tip structure \cite{Sebastian:09}, 
the methodology has not been used systematically to explore spacing under 
transient solidification conditions. Indeed the ability to model  cell, dendrite and 
seaweed structure, kinetic and surface tension anisotropy, different mobility, 
different thermal conditions and different initial condition makes phase field 
modeling an ideal theoretical test ground to explore transient spacing 
development and how it may relate to the steady structures. 

This paper reports new experiments that  study primary spacing selection in 
directionally solidified Al-Cu alloys cooled under transient conditions closely related 
to those encountered in strip casting of Al alloys.  The transient thermal gradient and 
interface speed are measured and correlated to measured dendrite spacing. Our 
results are shown to be inconsistent with steady state or transient geometric 
theories. Instead they are more consistent with stable ranges of spacings versus 
front speed, connected by rapid changes in spacing at particular interface velocities. 
We also present new two dimensional phase field simulations that 
exhibit the same behavior as our experimental data. Analysis of the phase field 
simulations is used to shed some light on the morphological development 
of dendrite arms during solidification under transient conditions. In order to manage 
the length of this paper, a sequel paper  to this one  will further explore the theoretical 
implications of the phase fields simulations presented here, focusing in particular, on 
the  theoretical connection between transient behavior reported in this paper to the steady 
state behavior previously reported.

\section{Experimental procedure}
\label{expt_methods}

\begin{figure}[!htp]
\begin{center}
\includegraphics[width=0.5\textwidth]{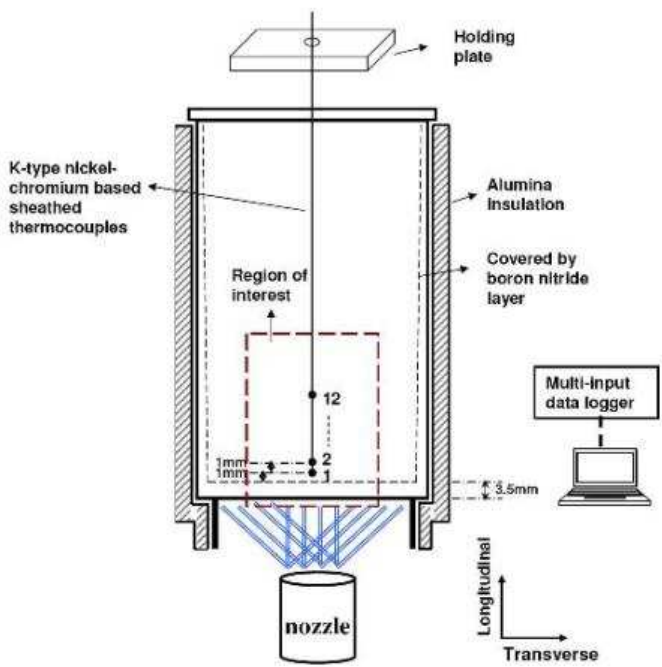}
\caption{Schematic of the set-up used to obtain upward directional solidification. \label{setup}}
\end{center}
\end{figure}

As-received Al-0.34wt\%Cu samples were used to study solidification 
microstructure evolution under transient cooling conditions. The experimental set-up 
is shown in figure~(\ref{setup}). It consists of a cylindrical stainless steel crucible that 
is water jet cooled from below to promote upward solidification. The crucible is shielded 
by a cylindrical alumina insulation with a thickness of about 10mm to prevent radial 
heat extraction. The inner, outer and bottom parts of the crucible were covered with 
a thin layer of sprayed boron nitride in order to reduce the heat extraction through the 
walls as well as providing a more uniform chilling surface at the bottom. The pressure 
of the spray was chosen high to  prevents the formation of bubbles at the water-chilling 
wall interface due to the local vaporization. 

The temperature is measured at different heights from the bottom with K-type nickel-chromium 
based sheathed thermocouples that are tightened along a plate and are 
inserted into the melt through the top of the crucible as shown in the figure. The diameter of 
the chilling surface is set to 50mm (i.e. the bottom plate) and the thickness of the chilling 
surface to 3.5mm. The large chilling diameter helps to reduce the influence of the 
thermocouples' diameter, about 1mm, on the solidification process. A set of thermocouples 
were aligned in the axial (vertical) direction starting at 1mm away from the chilling surface and 
separated from each other by 1mm. In order to ascertain the one dimensionality of the heat 
flow in the vertical direction, an additional thermocouple was positioned 12mm  radially
from the aligned thermocouples. 

The output from the thermocouples is acquired through a NI SCXI-1600 data logger 
and the calibration of which was set at the melting point of pure aluminum and pure zinc.
Before pouring the melt into the crucible, the crucible along with the surrounding insulation 
and the alloy were heated in the same furnace to a temperature of 1.1 times the liquidus 
temperature of the alloy, to compensate for the heat loss during the experiment. 

The cooling curves obtained from the thermocouples define a region of radially uniform 
temperature that varies essentially in the growth direction. This region is referred to as 
our region of interest. Outside this region, there is a temperature gradient towards the wall 
such that the unidirectional heat flow breaks down.  Far from the bottom chilling plate,  
outside the region of interest, the solidification microstructure mostly consists of equixed 
dendrites. Within our region of interest the solidification microstructure is columnar in nature
(i.e. oriented cells, dendrites or seaweed branches). We disregard any data outside our region 
of interest. As will be shown below, the grains examined within the region of interest are large 
enough in the direction transverse to the heat flow to disregard boundary effects. Only data 
from grains with transverse size larger than 1$mm$ are reported here.  In each sample 
solidified, three different directionally solidified grains emanating from the chill surface 
were analyzed for reproducibility and to provide statistical error bars we report in our results.

\begin{figure}[!htp]
\begin{center}
\includegraphics[width=0.4\textwidth]{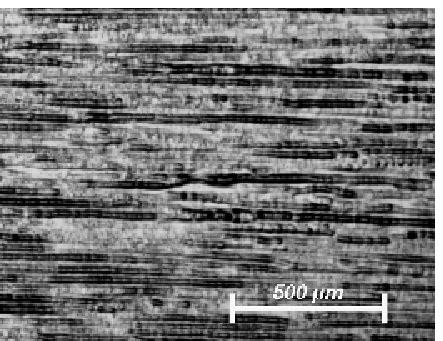}
\includegraphics[width=0.4\textwidth]{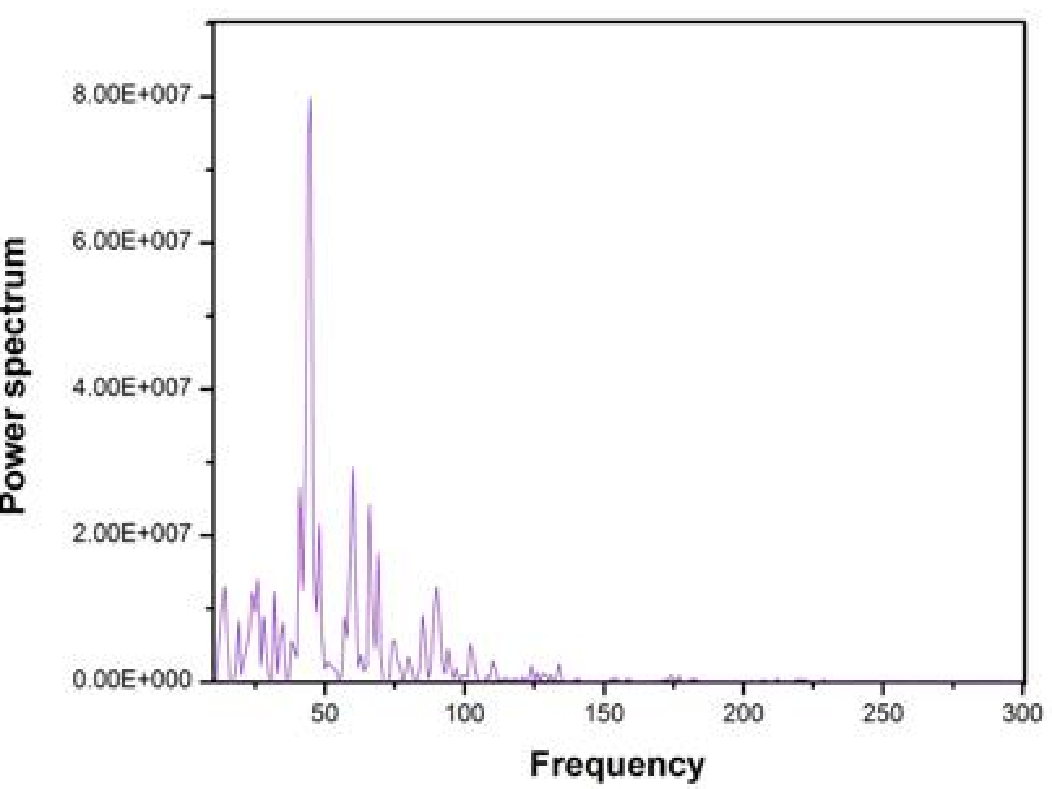}
\caption{(Top) cutaway of the longitudinal solidification microstructure, growing from left to right. 
(Bottom)  Unidimensional power spectrum  of a transverse cut extracted from the longitudinal 
microstructure at a position 13$mm$ from the chilling surface. \label{powerspectrum_1D}}
\end{center}
\end{figure}

\begin{figure}[!htp]
\begin{center}
\includegraphics[width=0.4\textwidth]{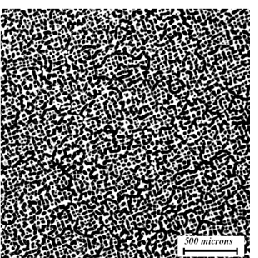}
\includegraphics[width=0.4\textwidth]{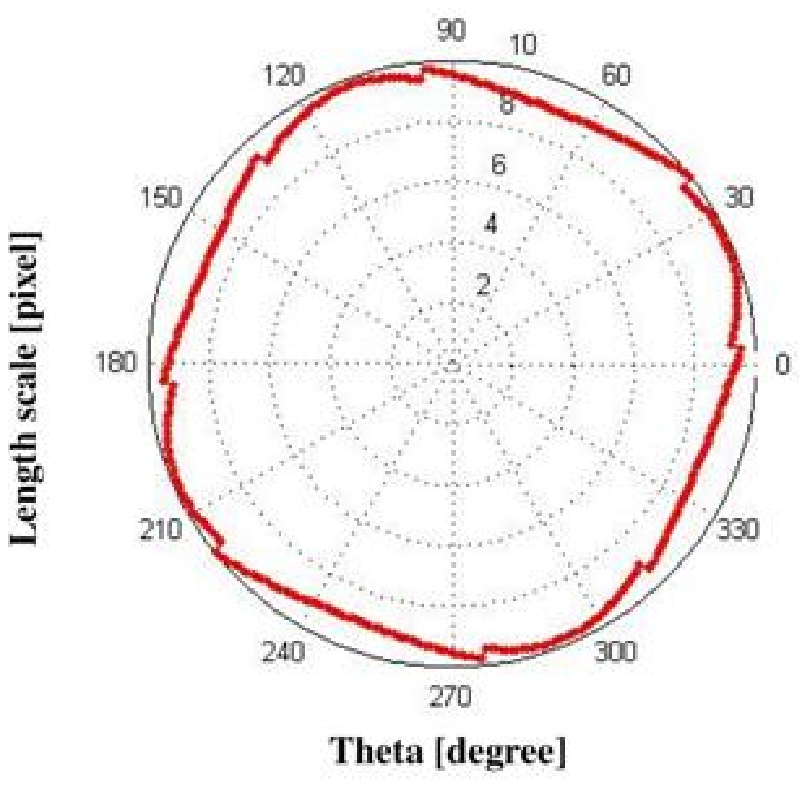}
\caption{(Top) Typical cross section of transverse dendrite microstructure. (Bottom) polar plot 
representing average length scale versus orientation in the transverse image in top frame. 
\label{powerspectrum_2D}}
\end{center}
\end{figure}

Figure~(\ref{powerspectrum_1D}) shows the longitudinal morphology of the dendrite 
microstructure cut out from a grain in one of our samples. The microstructure 
appears cellular in nature, while there is some evidence of seaweed-like tip-splitting 
throughout the image.  The microstructural length scale in the direction transverse to 
that of the heat flow (right to left in the figure) was analyzed at different distances 
from the chilling surface using power spectral analysis.  A typical power spectrum is 
shown in the bottom frame of figure~(\ref{powerspectrum_1D}).  The x-axis of the power 
spectrum denotes the frequency of the corresponding wave vector. The main peak is 
associated to the primary spacing and is consistent with would be obtained by the 
ASTM line intersection method. The smaller, high frequency,  peaks account for 
smaller wavelengths such as sidebranches,  splitting tips or  the discreteness of the 
pixels in an image. The smaller frequency peaks correspond potentially to  
longer-wavelength interactions between dendrites branches.
Figure~(\ref{powerspectrum_2D}) shows a typical image of the corresponding transverse
microstructure, cut away from a grain at distance 15$mm$ from the chilling surface. The 
bottom frame of figure~(\ref{powerspectrum_2D}) shows a polar plot showing the average 
length scale versus angle in the data in the top frame, obtained from a digitization of the
corresponding cross section in the top frame.   
The polar plot was obtained by averaging the two-point direct correlation function from 
data such as that in the top frame of figure~(\ref{powerspectrum_2D})  along lines of different orientation. 
The polar --or ``rose"--  plot in figure~(\ref{powerspectrum_2D}) represent the mean inter-branch 
spacing in the data along different orientations.  Details of the procedure by which the 
polar plot is constructed can be found in \cite{Kuchnio:09}.\\

\section{Phase Field Modeling}
\label{PF_mod}

Phase field simulations modeled the Al-Cu alloy system in the dilute limit of the phase 
diagram, which comprises straight solidus and liquidus lines of slopes of $m/k$ and 
$m$, respectively. The equilibrium interface concentration jump at any temperature is 
thus given by  partition relation $c_{s}=kc_{l}$, where $c_{s}$  ($c_{l}$) is the molar 
concentration of impurities at the solid (liquid) side of the interface, and $k$ is the 
partition coefficient.

Simulations formally neglected the latent heat by imposing the temperature field 
by the form  $T(z,t)=T_{0}+G(t)(z-z_{0}-\int_0^t V_{p}(t')dt')$, where $T(z_{0},0)=T_{0}$,  
a reference temperature, while $G(t)$ and $V_p(t)$ are the local thermal gradient 
and pulling speed, respectively. These were extracted from our experiments as 
discussed further below. We neglected convection in the liquid, as the transport 
of impurities becomes governed essentially entirely by diffusion \cite{Hongda:09}. 
Moreover, since diffusion of impurities in the solid is typically several orders of 
magnitude lower than in the liquid, diffusion in the solids is neglected. 

Under the above assumptions, mass conservation 
across the interface takes the form $c_{l}(1-k)v_{n}=-D\partial_{n}c|_{l}$, where $D$ is the 
solute diffusion coefficient in the liquid and $\partial_{n}|_{l}$ is the partial derivative in the 
direction normal to the interface, taken on the liquid side. The temperature at the interface, 
which is assumed to be in local equilibrium, is given by the Gibbs-Thomson relation 
$T=T_{m}-|m|c_{l}-\Gamma\kappa-v_{n}/\mu_{k}$, where $T_{m}$ is the melting temperature 
of the pure material, $\Gamma=\gamma T_{m}/L$ is the Gibbs-Thomson coefficient, $\gamma$ 
is the interfacial free energy, $L$ the latent heat of fusion per volume, $\kappa$ is the interface 
curvature, $v_{n}$ is the normal interface velocity and $\mu_{k}$ is the atomic mobility at 
the interface. 

The underlying crystalline structure that defines the anisotropy of surface tension 
(or interface mobility) is modeled by through a commonly used fourfold symmetry 
anisotropy function $a(\hat{n})=1-3\epsilon+4\epsilon(\hat{n}_{x}^{4}+\hat{n}_{y}^{4}+\hat{n}_{z}^{4})$, 
where $\epsilon$ is the anisotropy strength and $\hat{n}$ is the unit normal at the the 
interface. In two dimensions  this function can be re-written as 
$a(\theta)=1-\epsilon cos(4\theta)$, where $\theta$ is the angle between the normal 
direction to the interface and an underlying crystalline axis (e.g $<100>$ in a cubic crystal).

Taking as reference the concentration on the liquid side of the interface 
$c_{l}^{0}=c_{0}/k$ (where $c_0$ is the average alloy concentration), 
the following standard one-sided sharp interface  directional
solidification kinetics are modeled:
\bea
\partial_{t}c=D\nabla^{2}c \label{eq.m4} \\
c_{l}(1-k)v_{n}=-D\partial_{n}|^{+} \label{eq.m5} \\
c_{l}/c_{l}^{0}=1-(1-k)\kappa d_{0}a(\hat{n}) \nonumber \\
-(1-k)\left(z-\int_0^t V_{p}(t')dt' \right)/l_{T}-(1-k)\beta v_{n} \label{eq.m6}
\eea
where $d_{0}=\Gamma/\Delta T_{0}$ is the solutal capillarity length, 
$\Delta T_{0}=|m|(1-k)c_{l}^{0}$ the freezing range, $l_{T}=\Delta T_{0}/G$ 
the thermal length, and $\beta=1/(\mu_{k}\Delta T_{0})$ the kinetic coefficient.

The phase-field model employed to emulate Eqs.~(\ref{eq.m4})-(\ref{eq.m6}) 
is designed for quantitative simulations through the use of a thin interface analysis 
developed by Karma and co-workers \cite{Karma:01,Echebarria:04}. This 
analysis makes it possible to emulate a specified capillary length and kinetic 
coefficient to second order accuracy in the ratio of the interface width to capillary 
length. The details of this model have been presented and discussed at length in 
Refs.~\cite{Karma:01,Echebarria:04} and thus only a brief description is included 
below. A general review of the phase-field method can be found in 
\cite{Boettinger:02,review:05}. 

A scalar phase field parameter $\phi$ is employed, which
takes on a constant value in each phase and varies sharply but smoothly across a 
diffuse interface. The phase field is used to interpolate the free energy 
density and mobility  between the bulk phases. Its equation of motion  
guarantees the system evolves towards a minimum of the free energy 
of the system. We define a phase-field variable which takes the value 
$\phi=1 (\phi=-1)$ in the solid (liquid). The concentration  $c(\vec{x},t)$ is 
characterized by through a generalization of the field $\tilde{U}=(c-c_{l}^{0})/(c_{l}^{0}(1-k))$, 
which represents the local supersaturation with respect to the point $(c_{l}^{0},T_{0})$, 
measured in units of the equilibrium concentration gap at that temperature. This 
generalized supersaturation field is given by 
\bea
U=\frac{1}{1-k}\Big(\frac{c/c_{l}^{0}}{(1-\phi)/2+k(1+\phi)/2}-1\Big)
\label{eq.pf1}
\eea
In term of the fields $c$, $\phi$ and $U$, the phase-field model referred to 
above is given by
\bea
\tau(\hat{n})\Big(1-(1-k)\frac{(z-z_{\rm int})}{l_{T}}\Big)\frac{\partial\phi}{\partial t}&=& w_{0}^{2}\vec{\nabla}\big[a(\hat{n})^{2}\vec{\nabla}\phi\big] \nonumber \\
+\phi-\phi^{3}-\lambda (1-\phi^{2})^{2}\big(U&+&\frac{z-z_{\rm int}}{l_{T}}\big) \label{eq.pf21} \\
\Big(\frac{1+k}{2}-\frac{1-k}{2}\phi\Big)\frac{\partial U}{\partial t}&=&\vec{\nabla}\Bigg[q(\phi)D \vec{\nabla}U \label{eq.pf22} \\
-\alpha w_{0}\Big(1+(1-k)U\Big)\hat{n}\frac{\partial\phi} {\partial t} \Bigg]&+&\Big(\frac{1+(1-k)U}{2}\Big)\frac{\partial\phi}{\partial t}  \nonumber
\eea
where $z_{\rm int} \equiv \int_0^t V_{p}\, dt'$ is the interface position, 
$\hat{n} \equiv -(\vec{\nabla}\phi)/(|\vec{\nabla}\phi|)$ defines the unit vector normal 
to the interface, $\tau(\hat{n})=\tau_{0} \cdot a^{2}(\hat{n})$ is the phase-field 
orientation dependent relaxation time and $a(\hat{n})=1-3\epsilon+
4\epsilon[(\partial_{x}\phi)^{4}+(\partial_{z}\phi)^{4}]$ imposes a fourfold 
anisotropy with strength $\epsilon$ in two dimensions. The function $q(\phi)=(1-\phi)/2$ 
dictates how the diffusivity varies across the interface. The interface thickness is given 
by $w_{0}$ while $\lambda$ is treated as numerical convergence parameter of the model.  

The parameters $\lambda$, $w_o$ and $\tau_o$ can be shown to be inter-related through 
the thin interface relations developed in Refs.~\cite{Karma:01,Echebarria:04} to map the 
above phase field model onto Eqs.~(\ref{eq.m4})-(\ref{eq.m6}) . Specifically, once a particular 
lambda is chosen, the thin interface relations establish a unique choice of $w_0$ and $\tau_0$ such 
as to yield the same $d_o$ and $\beta$ in simulations. The aim is to choose a rather diffuse 
value of $w_0$ in order to expedite numerical efficiency. In this work, we assume the 
interface kinetics coefficient $\beta \approx 0$, to lowest order.  This ability to quantitatively 
model the same materials parameters $\beta$ and $d_0$ is largely due to the term containing the constant 
$\alpha$ in Eq.~(\ref{eq.pf22}). The term is called the so-called ``antitrapping current", whose 
function is to self-consistently  counter the spurious effects of an interface thickness that is artificially 
enlarged for practical purposes. 

The material parameters employed represent an Al-Cu alloy and are presented in 
table \ref{table.materialparam}.
\begin{table}[!htp]
\begin{center}
\begin {tabular}{|c|c|}
\hline
$|m|$ ($K/wt\%$) & $3.00$ \\
\hline
$c_{0}$ ($wt\%$) & $0.34$ \\
\hline
$k$ & $0.15$\\
\hline
$D$ ($\mu m^2/s$) & $3400$ \\
\hline
$\Gamma$ ($K \cdot \mu m$) & $0.10$\\
\hline
$\epsilon$ & $0.02$\\
\hline
\end {tabular}
\end{center} 
\caption{Material parameters that define the samples employed. $m$ is the 
liquidus slope, $c_{0}$ the alloy composition, $k$ the partition coefficient, 
$D$ the diffusivity of impurities in the liquid, $\Gamma$ the Gibbs-Thomson 
constant and $\epsilon$ the anisotropy strength. \label{table.materialparam}}
\end{table}
We solve the phase field equations using either a finite difference explicit Euler scheme 
on a uniform mesh, or, for simulations on larger scale, with a new finite difference adaptive 
mesh algorithm that utilizes a data structure developed Provatas and co-workers 
\cite{Provatas:98,Athreya:07}. 

\begin{figure}[!htp]
\begin{center}
\includegraphics[width=0.4\textwidth,height=0.22\textheight]{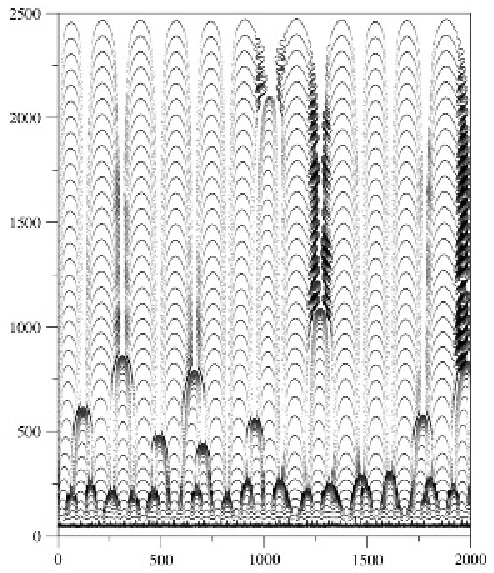}
\includegraphics[width=0.4\textwidth,height=0.22\textheight]{fft}
\caption{(Top) Simulated evolution of a directionally solidified dendrite array grown under 
constant velocity and thermal gradient. Interface initially morphologically noisy.  Distances are 
in $\mu m$.  (Bottom) Spectral analysis of the last recorded interface structure (black line) as 
well as that of a close earlier time (red line).\label{fft}}
\end{center}
\end{figure}

Figure~(\ref{fft}) shows an example of a typical sequence of directionally solidified 
dendrite arms growth under steady state conditions, i.e. a constant thermal gradient  
($G=5~K/mm$) and puling speed ($V_p=10~\mu m/sec$). In order to systematically 
study the evolution of interface structure, and, in particular,  the selection of the 
columnar microstructure spacing, we also perform a power spectral analysis on 
simulated interface profiles using a Fast Fourier transform. An example of such 
evolution is also shown in figure~(\ref{fft}). It is noted that  the emergence of a split 
in the main peak of the spectrum at the earlier stage reveals the interaction between 
dendrites that eventually leads to cell elimination.

The phase field model was primarily used to simulate transient 
cooling conditions relevant to the experimental situation described in 
section~(\ref{expt_methods}). This was done by using the thermocouple 
data to extract the local thermal gradient across the solid-liquid interface 
and the effective velocity of the solidification front, which were then fitted 
to provide functions used to determine $G(t)$ and $V_{p}(t)$. The pulling 
speed was modeled after a fit of the front velocity obtained from experiments 
which, given that the interface is initially positioned at $T_{L}$, is 
systematically lower than the actual front velocity, the discrepancy decreasing 
as the system evolves. Further details about this are discussed in 
section~(\ref{spacing_results_sim}).

\begin{figure}[!htp]
\begin{center}
\includegraphics[width=0.35\textwidth,height=0.22\textheight]{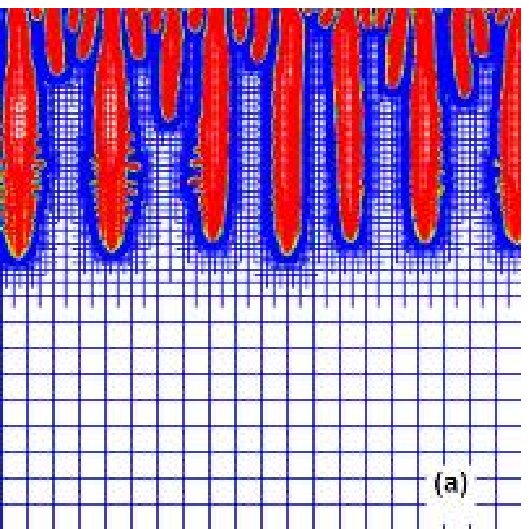}
\includegraphics[width=0.35\textwidth,height=0.22\textheight]{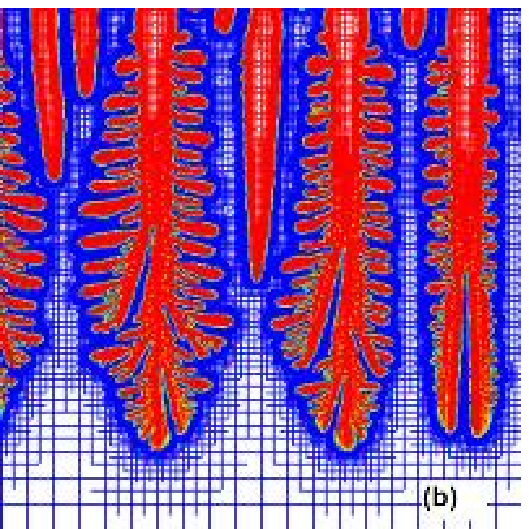}
\includegraphics[width=0.35\textwidth,height=0.22\textheight]{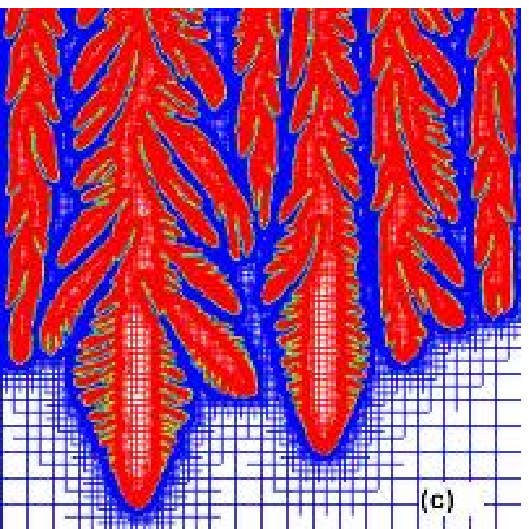}
\includegraphics[width=0.35\textwidth,height=0.22\textheight]{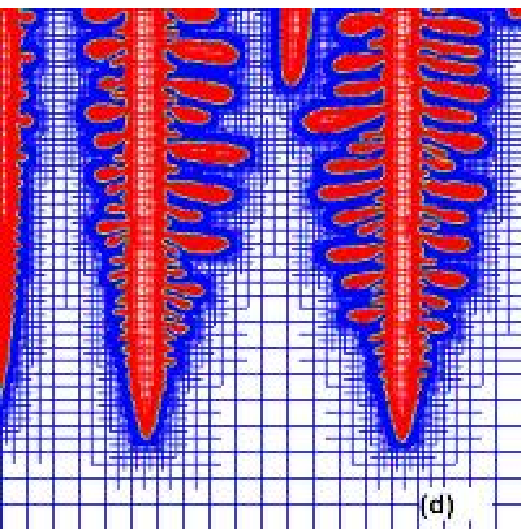}
\caption{Dendritic microstructure evolved during the PF simulation utilizing adaptive 
mesh refinement at the positions(a) 600 (b) 1200 (c) 3600 and (d) 11400 microns from 
the initial position of the interface, corresponding to $T_{L}$. Each image is a zoom-in
of the interface and is 65 microns in dimension. The colour represents concentration.
\label{simulation}}
\end{center}
\end{figure}

Figure~(\ref{simulation}) shows a typical spatiotemporal evolution of 
dendritic microstructures from our transient dynamic simulations. The 
morphological evolution has features in common with the experimental 
morphology in figure~(\ref{powerspectrum_1D}). Specifically, at early 
times,  when the velocity is fastest, the morphology is cellular and then 
starts to exhibit a series of kinetically-induced tip-splitting instabilities, 
giving rise to rising columns of seaweed-like structures. At later times, 
when the velocity decreases, the cellular branches emerge once more.  
The analysis of our phase field simulations and the experimental data 
of section~(\ref{expt_methods}) is discussed below.

\section{Results and Discussion}

\subsection{Finite Size Effects} 

\begin{figure}[!htp]
\begin{center}
\includegraphics[width=0.45\textwidth,height=0.35\textheight]{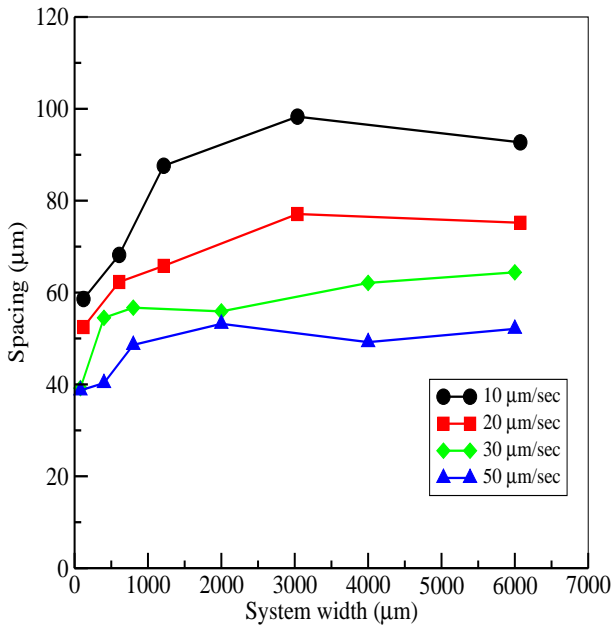}
\caption{Steady-state spacing versus transverse sample width for different 
pulling speeds. The effect of the boundary becomes less prominent as the 
sample width increases. G=$5$ $K/mm$. \label{systemsize}}
\end{center}
\end{figure}

In order to avoid boundary effects on the dendrite arm spacing, we 
study arm spacing in as large a single grain as possible. The simulations, on the other hand,
are more time consuming for larger systems. To estimate a convinient grain 
(or system) size to use experimentally and theoretically in our spacing selection analysis, 
we studied the dependence of dendrite spacing on the system 
size in phase field simulations with constant control parameters.
Figure~(\ref{systemsize}) shows the final steady state spacing as a function of transverse 
system size (i.e. grain size) for four different puling speeds. In all cases we started with a 
morphologically noisy interface.

These results indicate boundary effects become negligible at systems (grain sizes) larger 
than about $10^3$ $\mu m$, even for the smallest rate of solidification, which leads  to the
largest spacing. The solidification rates studied experimentally and theoretically 
in the transient solidification analysis below are higher than those studied here, making 
finite size effects even more negligible in systems of $10^3$ $\mu m$ or larger.

\subsection{Dendrite spacing evolution: experiments}
\label{spacing_results}

\begin{figure}[!htp]
\begin{center}
\includegraphics[width=0.48\textwidth,height=0.35\textheight]{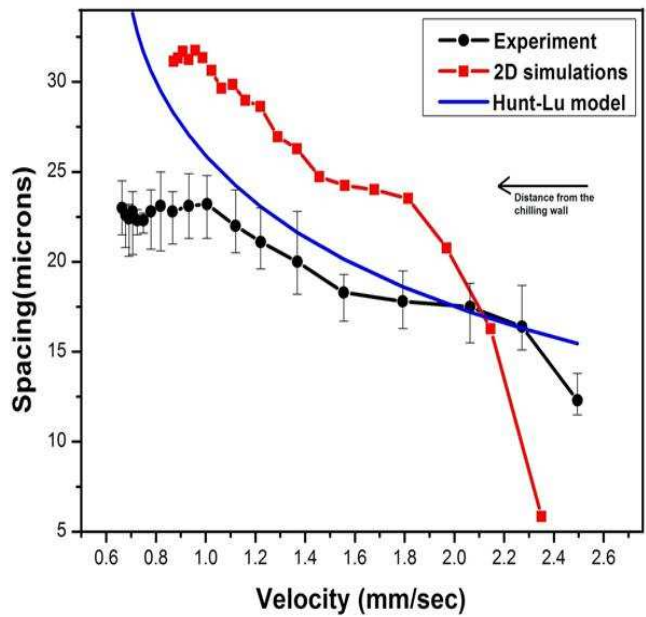}
\caption{Spacing as a function of velocity obtained experimentally (black curve) and numerically (red curve). The blue curve corresponds to the relationship obtained by Hunt and Lu under unsteady state solidification. (Note that G(t) is also dynamically changing at each point.) \label{spacingvelocity}}
\end{center}
\end{figure}

Figure~(\ref{spacingvelocity}) shows experimental (black curve) and simulated (red curve) 
plots of the transverse dendrite spacing as a function of the front velocity, which varies 
during solidification. Also shown for comparison is the corresponding 
primary arm spacing determined from the unsteady state model of Hunt and Lu \cite{Hunt-Lu:96} 
(blue curve). In the experiments, temperature gradient is coupled with the velocity and are not 
independent. The  solidification rate is obtained from the cooling curves by considering 
that a thermocouple registers a sharp change in the temperature slope (with respect to time)
when the solidification front passes through it.  The experimental spacing shown is that obtained 
from transverse sections. Analysis of the longitudinal sections shows the same qualitative behavior, 
although the values are different, as expected. It is recalled that the experimental spacing data 
is obtained from different grains of the same experiment.

The experimental results in figure~(\ref{spacingvelocity}) are consistent with those reported  
by Losert et al \cite{Losert:96}. Namely, the spacing exhibits regions of very slow change, 
between which it changes rapidly.  It is noteworthy that the experimental data does not compare
well quantitatively and, especially, qualitatively  with the unsteady state model of Hunt and Lu. 
Plausible reasons for this will be addressed in section~(\ref{discussion}).

\begin{figure}[!htp]
\begin{center}
\includegraphics[width=0.5\textwidth]{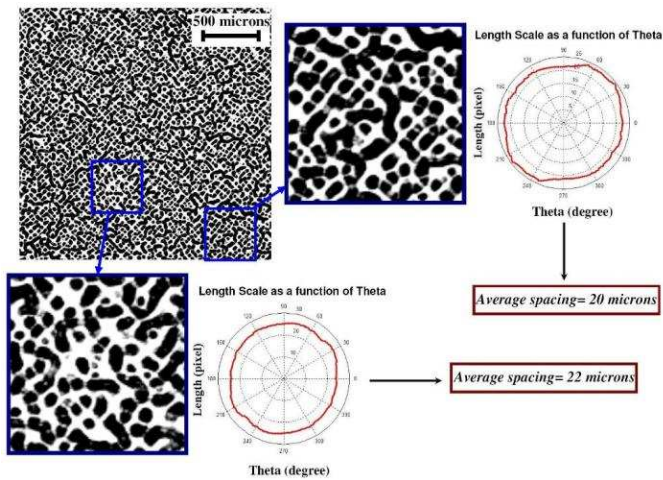}
\caption{separate zones within a grain at the same height level from the chilling 
wall which exhibit different values of average spacing.\label{domain}}
\end{center}
\end{figure}

Losert et al associate a rapid change or jump in spacing with the period doubling instability 
studied by Warren and Langer, citing boundary effects to account for the discrepancy of the 
jump being less than a factor of two. We propose that the discrepancy is also due to the fact that 
instabilities do not affect the system as a whole but are confined to smaller domains. The 
solute field ahead of the interface is not periodic and a localized disturbance has to propagate 
along the front before affecting other regions. Hence there is a delay between an instability 
in one region and its consequences to farther away dendrites. The spacing at any given time is 
thus the average of spacing in different domains, rather than a consequence of an instability 
affecting the system instantaneously. We examined this hypothesis in our experiments 
by examining different regions of a single transverse section at a specific position away from 
the chilling surface.  Regions with different average spacing are found, consistent with our 
hypothesis. Figure~(\ref{domain}) shows two separated regions within same transverse 
section of a grain which have different average spacing.


\subsection{Dendrite spacing evolution: simulations}
\label{spacing_results_sim}

In order to approximate the growth conditions similar to those in our experiments, 
our directional solidification  simulations included a variable thermal gradient and 
pulling speed, the values of which were set by fitting the corresponding curves 
obtained from experiments. Of-course, the pulling speed is not the same as the 
front velocity, especially under transient effects, since the interface moves within 
the coexistence region as growth conditions vary. This discrepancy is largest at 
the onset of the simulation due to the initial conditions.  

The two dimensional simulations in figure~(\ref{spacingvelocity}) show remarkably similar  
behavior to the experiments,  consistent with the idea that the branch spacings are 
confined to a band of slowly varying spacing separated by rapid change at particular 
velocities. The quantitative discrepancies between experiments and numerical results 
are to be expected due to the different growth conditions of the simulations compared 
to the experiments and that the simulations are two dimensional.

\begin{figure}[!htp]
\begin{center}
\includegraphics[width=0.4\textwidth]{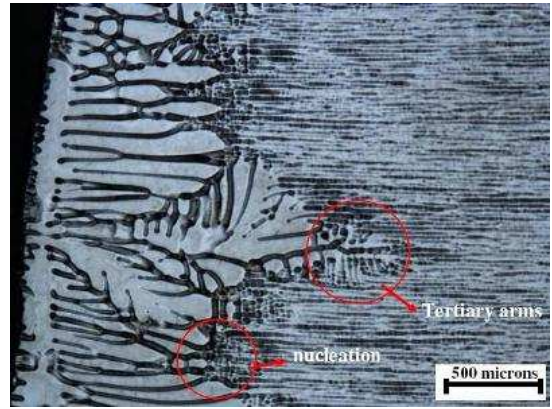}
\caption{Cast Al-Cu alloy showcasing the sharp change in morphology and spacing that 
occurs during a transition from low cooling rate (large-scale structure) to rapid cooling rate (finer
structure).
\label{rapidcooling}}
\end{center}
\end{figure}

It is noteworthy that the mean peak of the power spectra of the simulation data essentially 
captures the main branch spacing and not the seaweed-like sub-structure evidenced in
figure~(\ref{simulation}). The latter structures are present at early and intermediate times 
and likely arise due to interface kinetics induced by the high solidification rate. As mentioned 
above, these structures appear in the experimental data as well (see figure~\ref{powerspectrum_1D}). 
To illustrate this further, we cooled a sample at very low rate for a short period of time an then 
increased the cooling rate abruptly. The resulting microstructure is shown in figure~(\ref{rapidcooling}). 
At low cooling rate,  where the microstructure is larger,  the microstructure comprises both 
cellular and seaweed-like tip splitting and side branching. After the cooling rate is increased, 
a finer structure emerges from the tips of these tertiary branches as well as nucleation events.  

\section{Mechanisms for Spacing Evolution}
\label{discussion}
 
The initial pulling speed implies the diffusion length is several orders of magnitude 
smaller than the system width. Moreover, even though the pulling speed decreases 
with time, the diffusion length stays several orders of magnitude below the system 
size. At the same time, the diffusion length is within an order of magnitude from the 
dendritic spacing. Therefore, each dendrite branch can be  considered to be in  
one of many,  weakly interacting, sub-arrays of branches. The change in cooling 
conditions, on the other hand, is a global event that affects the whole system at the 
same time. It is therefore to be expected that the quasi-independent sub-arrays of 
branches will react differently to the globally changing cooling conditions. As a 
result, each sub-array may undergo instabilities --to some extent-- independent 
from each other. This will influence the "jumps" in the spacing versus velocity 
curve. The rate of change in the cooling conditions may also play an important role, 
as the change in the global cooling conditions may cause the solute field associate 
with different regions to transfer information to neighbouring regions at different times. 

For a small number of dendrites branches in a system, the larger the change in 
the spacing due to a given branch elimination event. Conversely, a large enough 
system, or smaller the spacing, should exhibit a {\it smoother} evolution of spacing. 
For the transient conditions we explored, there appears to exist quasi-independent 
domains which undergo instabilities without, for the most part, affecting each other. 
After an instability, such as cell elimination, affects one, or several, entire domains, 
the average interdendritic spacing of the entire system will experience a sharp change. 
On the other hand, we also expect periods of low latency in which the overall 
spacing does not change, or evolves very slowly, while the different regions slowly interact 
through diffusion of the solute field. These observations are consistent with our findings 
and also with those of Losert et al \cite {Losert:96}, in which he changed the pulling 
speed in steps and found a sharp jump in the spacing after it went almost unchanged 
for a range of pulling speeds.

\begin{figure}[!htp]
\begin{center}
\includegraphics[width=0.4\textwidth]{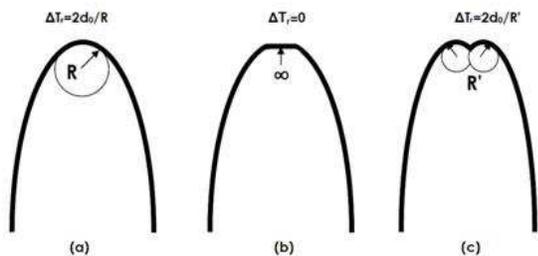}
\caption{Schematic representation of different stages during the tip splitting mechanism. 
In the flattening stage (b), the undercooling due to the Gibbs Thomson effect is essentially 
zero.\label{tipsplitting}}
\end{center}
\end{figure}

The existence of plateaus connected by rapid changes in spacing points to a dynamics 
in which an energy barrier has to be overcome for the spacing to adapt. Specifically,  
for the tip of a cell or branch to split, it effectively needs to pass through a flattening stage in 
which the tip radius becomes effectively infinite.  This lowers the interface undercooling 
(i.e. the contribution of dendrite tip curvature to interface undercooling, $d_{0}\kappa$ , 
becomes zero), preventing tip splitting until a larger local interface velocity is reached. 
Increasing the velocity reduces the diffusion length and shortens the distance over which
solute is rejected. This effect acts to reduce arm spacing. Hence, there will be a competition 
between the two effects, keeping the spacing constant due to the first effect until the 
driving force provided by second phenomenon is large enough to force dendrites or 
cells to split and reduce their spacing. Figure~(\ref{tipsplitting}) schematically represents 
the stages of tip splitting.

\begin{figure}[!htp]
\begin{center}
\includegraphics[width=0.5\textwidth]{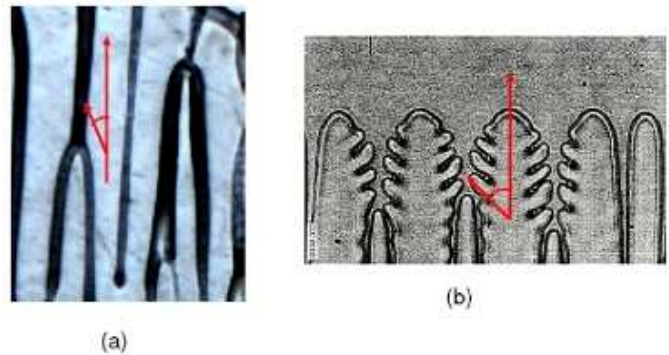}
\caption{Cell elimination caused by the neighbouring (a) primary arms or (b) secondary arms \cite{privateLosert} at high and low velocity solidification respectively.\label{cellelimination}}
\end{center}
\end{figure}

An increase in branch spacing via cell elimination occurs when a dendrite is 
blocked by either the secondary arms (in low speed solidification) of an 
adjacent dendrite or by the main stalk (in high speed solidification) of an 
adjacent dendrite. As shown in figure~(\ref{cellelimination}), in both cases, 
the seizing arm grows at an angle with a velocity component perpendicular to 
the growth direction of the primary arms (which is the direction of lowest energy, 
obtained as the product of anisotropy direction and heat flow direction). Unlike 
equiaxed growth, where secondary arms grow at a rate comparable to the growth 
rate of primary arms, in directional solidification the growth rate of secondary branches 
is negligible compared to that of the main trunks. Thus, the seizing mechanism 
described above is not able to act until diffusion-mediated interactions through 
the melt become significant.  As growth velocity decreases, solute diffuses a 
longer distance and if the diffusion length of a dendrite is large enough, it interacts 
with adjacent diffusion fields. It is at this point that a jump in system energy can occur. 

Hence, in the case of cell elimination, there is an energy increase due to the growth 
in any of the undesired directions and an energy decrease for changes that increase 
the distance between dendrites (i.e. characterized by non-interacting diffusion fields). 
These two phenomena dynamically compete, with the spacing change eventually 
determined by the dominant effect, where the former effect acts as a barrier against 
the change in the spacing and the latter provides the driving force for cell elimination. 
It is also plausible that over a small range of  cooling conditions (i.e. solidification rate, 
temperature gradient) these effects may balance each other, causing the spacing not to 
change very much, at least over some long-lived transient time (i.e. the plateau regions 
in the data).  Furthermore, the farther from steady state spacing the initial condition of 
the system, the higher the energy and the larger the driving force required for the system 
to perform branch  elimination. Thus, it is possible that different spacings can also exist at 
the same cooling conditions depending on the history of the system. This matter will be further
investigated in future work.

\section{Acknowledgments}
We would like to thank  the National Science and Engineering
Research Council of Canada (NSERC) and Novelis Inc 
for financial support of this work.

\end{document}